\documentclass{ws-rv9x6}
\usepackage{ws-rv-van}             % numbered citation/references (default)
\makeindex
\begin{document}

\chapter[Computational modelling of evolutionary systems]{Computational
modelling of  evolution: ecosystems and language \label{ch1}}

\author[A. Lipowski and D. Lipowska]{A. Lipowski$^{1)}$ and D. Lipowska$^{2)}$}
%\index[aindx]{Author, F.} % or \aindx{Author, F.}
%\index[aindx]{Author, S.} % or \aindx{Author, S.}

\address{$^{1)}$Faculty of Physics, Adam Mickiewicz University,\\
61-614 Pozna\'n, Poland,\\ e-mail: lipowski@amu.edu.pl}

\address{$^{2)}$Institute of Linguistics,Adam Mickiewicz University,\\
Pozna\'n, Poland, \\ email: lipowska@amu.edu.pl}
\begin{abstract}
Recently, computational modelling became a very important research
tool that enables us to study problems that for decades evaded
scientific analysis. Evolutionary systems are certainly examples
of such problems: they are composed of many units that might
reproduce, diffuse, mutate, die, or in some cases for example
communicate. These processes might be of some adaptive value, they
influence each other and occur on various time scales. That is why
such systems are so difficult to study. In this paper we briefly
review some computational approaches, as well as our
contributions, to the evolution of ecosystems and language. We
start from Lotka-Volterra equations and the modelling of simple
two-species prey-predator systems. Such systems are canonical
example for studying oscillatory behaviour in competitive
populations. Then we describe various approaches to study
long-term evolution of multi-species ecosystems. We emphasize the
need to use models that take into account both ecological and
evolutionary processes. Recently we introduced a simple model of
this kind, and its behaviour is briefly summarized. In this
multi-species prey-predator system, competition of predators for
preys and space results in evolutionary cycling. We suggest that
such a behaviour of the model might correspond to long-term
periodic changes of the biodiversity of the Earth ecosystem as
predicted by Raup and Sepkoski. Finally, we address the problem of
the emergence and development of language. It is becoming more and
more evident that any theory of language origin and development
must be consistent with darwinian principles of evolution.
Consequently, a number of techniques developed for modelling
evolution of complex ecosystems are being applied to the problem
of language. We briefly review some of these approaches. We also
discuss the behaviour of a recently introduced evolutionary
version of the naming-game model. In this model communicating
agents reach linguistic coherence via a bio-linguistic transition
which is due to the coupling of evolutionary and linguistic
abilities of agents.
\end{abstract}
\body
\section{Introduction}\label{intro}
Evolution is a fundamental property of life. It consists of two
basic and in a sense opposing ingredients. Mutations and
crossing-over are forces that increase variability of organisms
and eventually lead to the formation of new brands and species.
Against these processes acts selection and due to limited
resources only best adapted organisms survive and pass their
genetic material to the next generation.

Evolutionary forces are operating since the very early emergence
of life. To large extent they shaped the complicated pattern of
past and present speciation and extinction processes. Even
qualitative understanding of the dynamics that governs these
processes is a very challenging problem. Eldredge and
Gould\cite{ELDREDGE1972,SIMPSON1983} noted that palaeontological
data show that intensity of speciation and extinction processes
varied throughout the life history and periods of evolutionary
stagnation were interrupted with bursts of activity (punctuated
equilibrium). Such a pattern, at least for a physicist, resembles
the behaviour of a system at a critical point. Indeed, similarly
to critical systems, some palaeontological data can be also
described with power laws\cite{NEWMANPALMER2003}. Following the
work of Bak and Sneppen\cite{BAKSNEPP1993}, a lot of models that
try to explain why an Earth ecosystem might be considered as a
critical system, were examined\cite{NEWMANPALMER2003}. But the
quality of palaeontological data does not allow for definitive
statements and some alternative interpretations were also
proposed. In particular, an interesting conjecture was made by
Raup and Sepkoski who suggested that the pattern of bursts of
extinctions is actually periodic in time with a period of
approximately 26 mln. years\cite{RAUPSEPKOSKI1984}. Despite an
intensive research it is not clear what a factor could induce such
a periodicity, and certainly, it would be desirable to understand
the main macroevolutionary characteristics of Earth's ecosystem.

But it is not only intensity of extinction and speciation
processes that is of some interest. In the evolution of life one
can distinguish several radical changes that had a dramatic
consequences and lead to the emergence of new levels of
complexity\cite{MSMITH1997}. As examples of such changes
Maynard-Smith and Szathm\'ary list invention of genetic code,
transition from Prokaryotes to Eukaryotes, or appearance of
colonies. Certainly, such changes had a tremendous impact on the
evolution and the current state of the Earth's ecosystem. As a
last major transition of this kind, Maynard-Smith and Szathm\'ary
mention the emergence of human societies and language. To large
extent this transition was induced by cultural interactions. Such
interactions when coupled to evolutionary processes lead to the
immensely complex system - human society. According to
Dawkins\cite{DAWKINS1976}, cultural interactions open a new route
to the evolution that is no longer restricted to living forms.
Examples of such forms, called memes, include songs (not only
human), well known sentences and expressions, fashion or
architecture styles. As a matter of fact, events listed as major
transitions have an interesting feature in common - they created
new mechanisms of information transmission. Language certainly
enables transmission and storage of very complex cultural
information. Its emergence enormously speeded up the information
transfer between generations. Before that, for nearly four billion
years of life on Earth, the only information that could be used
for evolutionary purposes was encoded in the genome. With language
vast amount of information can be exchanged between humans and
passed on to subsequent generations. Most likely it was the
invention of language that enormously speeded up the the evolution
of our civilization and made humans in a sense a unique species.
Although the emergence and subsequent evolution of language had
tremendous influence, this process is still to large extent
mysterious and is considered as one of the most difficult problems
in science\cite{CHRISTIANSEN}.

Computer modelling seems to be a very promising technique to study
complex systems like ecosystems or langauge. In the present paper
we briefly review such an approach and present our results in this
field. In section~\ref{ibm} we briefly discuss population dynamics
of simple two-species prey-predator systems and classical
approaches in this field based on Lotka-Volterra equations. We
also argue that it is desirable  to use an alternative approach,
the so-called individual based modelling. An example of such a
model is described in section~\ref{lattice}. In this section we
discuss results of numerical simulations of the model concerning
especially the oscillatory behaviour.

Processes in simple ecosystems with constant number of
non-evolvable species (as described in section~\ref{ibm}) take
place on ecological time scale. To describe real, i.e., complex,
ecosystems we have to take into account also evolutionary
processes, such as speciations or extinctions. Such processes
operate on the so-called evolutionary time scale. Such a time
scale was usually regarded as much longer than ecological time
scale, however, there is a number of examples that show that they
are comparable\cite{THOMPSON,FUSSMANN}. In section~\ref{multi} we
briefly review models used to study complex ecosystems. In
particular, we emphasize the need to construct models that would
take into account both ecological and evolutionary processes. In
section~\ref{extin} we examine one of such models which is a
multi-species generalization of prey-predator model studied in
section~\ref{lattice}. Investigating extinction of species we show
that their intensity changes periodically in time. The period of
such oscillations is set by the mutation rate of the model. Since
evolutionary changes are rather slow, we expect that such
oscillations in the real ecosystem would have very long
periodicity. We suggest that such a behaviour agrees with the
conjecture of Raup and Sepkoski, but more detailed analysis of the
predictions of our model would be desirable. In the final part of
this section we suggest that our model might provide an insight
into a much different problem. Namely, we attempt to explain the
uniqueness of the coding mechanism of living cells as contrasted
with the multispecies structure of present-day ecosystems.
Apparently, at the early stage of life a primitive replicator
happened to invent the universal code that was so effective that
it spread over the entire ecosystem. However, at a certain point
such a single-species ecosystem become unstable and was replaced
by a multi-species ecosystem. In our model, upon changing a
control parameter, a similar transition (between single- and
multi-species ecosystems) takes place and we argue that it might
be analogous to the early-life transition.

In section~\ref{lang} we review computational studies on language
evolution. An important class of models is based on the so-called
naming game introduced by Steels\cite{STEELS1995}. Recently, we
examined an evolutionary version of this model and showed that
coupling of evolutionary and linguistic interactions leads to some
interesting effects\cite{LIPLIP2008}. Namely, for sufficiently
large intensity of linguistic interactions, there appears an
evolutionary pressure that rapidly increases linguistic abilities
and the model undergoes an abrupt bio-linguistic transition. In
such a way communicating agents establish a common vocabulary and
the model reaches the so-called linguistic coherence. Our model
incorporates both learning and evolution. Interaction of these two
factors, known as a Baldwin effect\cite{BALDWIN1896}, is recently
intensively studied also in the context of language
evolution\cite{YAMA2004}. Discussion of the Baldwin effect and
related properties in our model is also presented in
Section~\ref{lang}. We conclude in Section~\ref{concl}.
\section{Coarse-grained versus individual-based modelling of an ecosystem}\label{ibm}
Population dynamics provides the basis of the modelling of the
ecosystem. Pierre Verhulst, regarded as its founding father,
noticed that due to the finite environmental capacity the
unlimited growth of the population predicted by the linear growth
equation is unrealistic. Consequently, Verhulst proposed that the
time evolution of the population should be described by the
following equation\cite{VERHULST}
\begin{equation}
\frac{dx}{dt}=kx(1-x/x_M) \label{verh}
\end{equation}
 where $k$ is the growth rate, $x$ is the size of the population, and $x_M$
 is the environmental capacity.
 Equation (\ref{verh}), that is called logistic equation, found numerous
 applications in demographic studies.
 However, to describe any realistic ecosystem one should consider
 more, possibly interacting, populations. A step in this direction was
 made by Lotka who examined a simple autocatalytic reaction model\cite{LOTKA}.
 His work was followed by Volterra who wrote down essentially the
 same set of ordinary differential equations (ODE) studying the statistics of fish
 catches\cite{VOLTERRA}.
 Lotka-Volterra equations for two interacting populations of preys ($x$) and
 predators ($y$) can be written as
 \begin{eqnarray}
 \frac{dx}{dt} &=&x(a_1-a_2y)\nonumber \\
 \frac{dy}{dt} &=&y(-a_3+a_4x)
 \label{lv}
 \end{eqnarray}
 where $a_1,\ a_2,\ a_3$, and $a_4$ are some positive constants. Although eqs.
 (\ref{lv}) constitute the canonical model to study periodic oscillations
 in competitive systems\cite{MURRAY}, they were also criticized on various grounds.
 For example their
 solution depends on the initial condition, and the very form of eqs.~(\ref{lv})
 is structurally unstable. It means that their small modification
 (with e.g., higher order terms like $x^2y$)
 will typically destroy oscillatory behaviour.
Although there are some ODE models where such a limit cycle
behaviour is more stable~\cite{MAY72},
 an important feature of any realistic system is missing in eqs.~(\ref{verh}) or (\ref{lv}). Namely,
 they neglect
 spatial heterogenities. The simplest way to take them into account
 would be to consider $x$ and $y$ as spatially dependent quantities
 and then to replace eqs.~(\ref{verh}) or (\ref{lv}) with their
 partial differential analogs.  After such a modification
 eq.~(\ref{verh}) becomes the famous Fisher equation, that in the
 one-dimensional case has the form
\begin{equation}
\frac{dx}{dt}=kx(1-x/x_M)+D\frac{\partial^2 x}{\partial l^2}
\label{fisher}
\end{equation}
where $l$ is the spatial coordinate, $D$ is the diffusion constant
and $x=x(t,l)$ depends on $t$ and $l$. Various extensions of
(\ref{fisher}), that are called sometime reaction-diffusion
models, were also intensively studied in ecological
contexts\cite{HOLMES,MURRAY}.

Although description in terms of partial differential equations
takes into account some of spatial heterogenities, it is still
based on the coarse-grained quantities like $x(t,l)$ and that
means that it is essentially of the mean-field nature. Moreover,
kinetic coefficients ($k,\ D,\ x_M,\ a_1,\ a_2,\ a_3,\
a_4,\ldots$) that enter such equations are usually difficult to
determine from ecological data. Similar problems appear in
alternative approaches to spatially extended ecological models
based on coupled-map lattices\cite{HASSELL} or integrodifference
equations\cite{HARDIN}. It is thus worth to pursue an alternative
approach, the so-called individual based modelling, where to some
extent stochastic rules, mimicking realistic processes like death,
breeding or movement, are formulated at the level of individual
organisms. Models of ecosystems formulated within such an approach
are particularly suited for numerical computations and resemble
some nonequilibrium statistical mechanics models. Such a
similarity is very valuable since the behaviour of ecological
systems can be put in a wider perspective.
\section{Lattice prey-predator models}\label{lattice}
To simplify calculations individual-based models of prey-predator
systems are usually formulated on a cartesian $d$-dimensional
lattice of the linear size $N$. One can define dynamics of such
models in various ways, but to provide a detailed example we
present rules used in some of our previous
works\cite{LIP1999,LIPLIP2000}. In our model on each site $i$ of a
lattice there is a four-state variable $\epsilon_i=0,1,2,3$ which
corresponds to the site being empty ($\epsilon_i=0$), occupied by
a prey ($\epsilon_i=1$), occupied by a predator ($\epsilon_i=2$)
or occupied by a prey and a predator ($\epsilon_i=3$). Its
dynamics has one control parameter $r$ ($0\leq r \leq 1$) and is
specified as follows: \vspace{-3mm}
\begin{itemize}
\item Choose a site at random. \item With the probability $r$
update a prey at the chosen site, provided that there is one
(i.e., $\epsilon=1$ or 3); otherwise do nothing. Provided that at
least one neighbor of the chosen site is not occupied by a prey
(i.e., $\epsilon=0$ or 2), the prey (which is to be updated)
produces one offspring and places it on the empty neighboring site
(if there are more empty sites, one of them is chosen randomly).
Otherwise (i.e., when there is a prey on each neighboring site)
the prey does not breed (due to overcrowding). \item With the
probability $1-r$ update a predator at the chosen site, provided
that there is one (i.e., $\epsilon=2$ or 3). Provided that the
chosen site is occupied by a predator but is not occupied by a
prey ($\epsilon=2$), the predator dies (of hunger). If there is a
prey on that site (i.e., $\epsilon=3$), the predator survives and
consumes the prey from the site it occupies. If there is at least
one neighboring site which is not occupied by a predator, the
predator produces one offspring and places it on the empty site
(chosen randomly when there are more such sites).
\end{itemize}
\vspace{-3mm} As neighboring sites, i.e., sites where offsprings
can be placed, we usually consider the nearest neighbours, but
taking into account further neighbours does not change the results
qualitatively\cite{KOWALIK}. To characterize the behaviour of the
model let us introduce the densities of preys ($x$) and predators
($y$) defined as
\begin{equation}
x=\frac{1}{N^d}\sum_i
(\delta_{\epsilon_i,1}+\delta_{\epsilon_i,3}),\ \
y=\frac{1}{N^d}\sum_i
(\delta_{\epsilon_i,2}+\delta_{\epsilon_i,3}),
 \label{defxy} \end{equation}
where summation is over all $N^d$ sites $i$ and $\delta$ is
Kronecker's $\delta$-function. Of the main interest are actually
averages $\langle x \rangle$ and $\langle y \rangle$, where
averaging is over simulation time.

\begin{figure}
\vspace{3cm} \centerline{\psfig{file=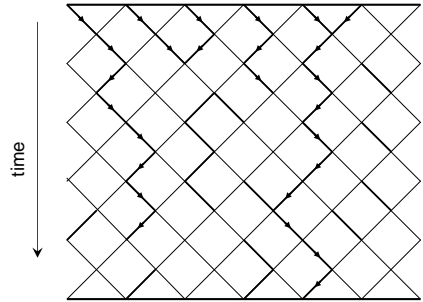,width=11cm,angle=0}}
\vspace{-6cm}\caption{In the directed percolation problem a
fraction of bonds on a lattice is permeable (thick lines). From
the top horizontal line water starts to flow downward through
permeable bonds. Bonds that are reachable by water are shown with
arrows. If at a certain (horizontal) level there would be no
water, there would be no water below that level (absorbing state).
When concentration of permeable bonds exceeds a certain threshold
value a cluster of permeable bonds that spans from top to bottom
is formed. At the threshold value the model is critical. A lot of
dynamical models with a single absorbing state belong to the
directed percolation universality class} \label{dp}
\end{figure}

From the above rules it follows that the model has two absorbing
states (i.e., once the model enter such a state it remains trapped
there for ever). The first one is filled with preys only ($x=1,\
y=0$) and the second one is empty ($x=0,\ y=0$).
Simulations\cite{LIP1999,LIPLIP2000,KOWALIK} show that for large
enough $r$, both populations coexist and the model is in the
active phase ($x>0,\ y>0$). When the update rate of preys $r$
decreases, their number becomes to small to support predators. For
sufficiently small $r$ predators die out and the model quickly
reaches the absorbing state where it is filled with preys (the
empty absorbing state has a negligible probability of being
reached by the model dynamics). The phase transition between
active and absorbing phase was observed at positive $r$ for
$d=1,2$ and 3. At least for $d=1$ (linear chain) Monte Carlo
simulations clearly show that the phase transition belongs to the
directed percolation (DP) universality class\cite{LIPLIP2000} (see
Fig.~\ref{dp}).
%%%%%%%%%%%%%%%%%%%%%%%%%%%%%%%%%%%%%%%%%%%%%%%%
%%%%%%%%%%%%%%%%%%%%%%%%%%%%%%%%%%%%%%%%%%%%%%%%

Such a behaviour is not surprising. There are by now convincing
numerical and analytical arguments that various models possessing
a single absorbing state generically belong to the DP universality
class~\cite{HINRICHSEN}. Moreover, models with multiple, but
asymmetric absorbing
 states (such as e.g., the model analysed in this section)
 also belong to this universality class (models with multiple but symmetric absorbing states typically belong
 to some other universality classes, or undergo discontinuous phase transitions~\cite{LIPDROZ}).
 However, studying the critical behaviour of models with absorbing-state phase transitions is
 not entirely straightforward. For finite systems (which is obviously the case in various
 simulational
 techniques) and close to the critical point, the model has a
 non-negligible probability of entering an absorbing state even
 when control parameters are such that the infinite system would
 remain in the active phase. Such a behaviour sets a size-dependent timescale (i.e., the lifetime of the active state)
 that severely affects simulations. A special technique, the so-called dynamical Monte Carlo, is
needed to obtain precise estimation of critical exponents for
models of this kind~\cite{GRASSBERGER,HINRICHSEN}.

Of our main interest, however, is the oscillatory behaviour of the
model. To examine it, we measured the variances of the densities
$x$ and $y$ as well as their Fourier transforms. Simulations show
that for $d=1$ and $d=2$ in the limit $N\rightarrow\infty$
stochastic fluctuations wash out the oscillatory behaviour and the
variances of densities vanish. However, for $d=3$ in the active
phase and close to the absorbing transition, there is a range of
$r$ where oscillatory behaviour survives in the limit
$N\rightarrow\infty$.\footnote{Such a conclusion is based on the
non-vanishing of variances of densities in this limit. Strictly
speaking, based solely on such a behaviour, one cannot exclude
that this is e.g., chaotic behaviour that sets in. However, a
pronounced peak in the Fourier transform of the time-dependent
densities strongly supports the oscillatory interpretation of the
numerical data.} Oscillations occur essentially for any initial
conditions and their period only weakly depends on the parameter
$r$.

\begin{figure}
\centerline{\psfig{file=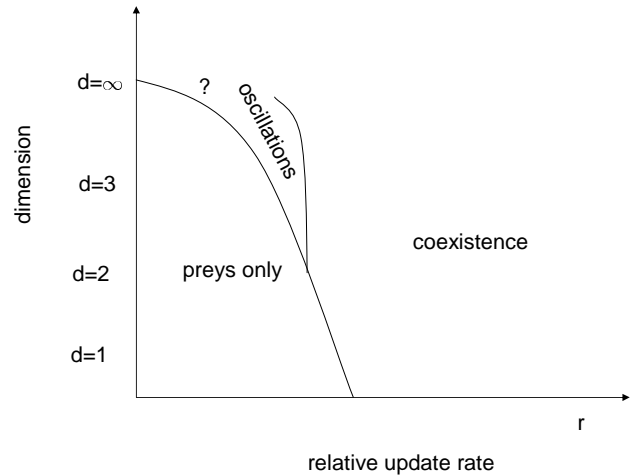,width=9cm,angle=0}}
\vspace{-5cm}\caption{The schematic phase diagram of the lattice
prey-predator model. The phase transition separating preys-only
and coexistence phases most likely belongs to the directed
percolation universality class\cite{LIPLIP2000}, but in more
general models other transitions are possible\cite{MOBILIA}. It is
not clear what is the fate of the oscillatory phase in $d>3$ case,
since the mean-field equations (\ref{mfa1}) (that most likely
correctly describe the model in large dimension) do not predict
the oscillatory regime.} \label{diagram}
\end{figure}

It is the dimension of the lattice $d$ that most likely plays an
important role. Indeed, simulations show that for $d=2$ models but
with larger number of neighbouring sites oscillations are again
washed out in the limit $N\rightarrow\infty$\cite{KOWALIK}. Such a
result is in agreement with some arguments of Grinstein \emph{et
al.}\cite{GRINSTEIN} who related temporal periodic phases of noisy
extended systems and smooth interfaces in growth models and
concluded that oscillations might exist but only for $d>d_c=2$.
The $(r,d)$ phase diagram of our model is sketched in
Fig.\ref{diagram}.

%%%%%%%%%%%%%%%%%%%%%%%%%%%%%%%%%%%%%%%%%%%%%%%%
%%%%%%%%%%%%%%%%%%%%%%%%%%%%%%%%%%%%%%%%%%%%%%%%

To get an additional insight into the behaviour of the model we
can write mean-field equations that describe the time evolution of
the densities $x$ and $y$. Simple arguments\cite{LIP1999} lead to
the following set of equations
\begin{eqnarray}
\frac{dx}{dt} & = & rx(1-x^w)-(1-r)xy \nonumber \\
\frac{dy}{dt} & = & (1-r)xy(1-y^w)-(1-r)y(1-x) \label{mfa1}
\end{eqnarray}
where $w$ is the number of neighboring sites, as defined in the
dynamical rules of the model (in most of our simulations
neighbouring sites were nearest neighbours and in such a case
$w=2d$). For example in the first equation of (\ref{mfa1}), the
first term ($rx(1-x^w)$) describes the growth rate of preys due to
updating a site with prey ($rx$) that happen to have at least one
empty neighbouring site ($(1-x^w)$). The second term ($(1-r)xy$)
describes the decrease rate of preys due to an update of a site
that happened to be a predator and that is also occupied by a
prey. However, predictions of approximation (\ref{mfa1}) even
qualitatively disagree with numerical simulations. In particular,
in any dimension $d$ the approximation (\ref{mfa1}) predicts that
for any positive $r$ there is no phase transition between active
and absorbing phases. Moreover, within this approach there is no
indication of the oscillatory phase, as observed in Monte Carlo
simulations for $d=3$.

In the approximation (\ref{mfa1}) the probability that a site is
occupied by a prey and predator is given as $xy$. This is of
course only an approximation, and a much better scheme is obtained
where this probability is considered as yet another variable
($z$), whose evolution follows from the dynamical rules of the
model. In such a way we arrive at the following set of
equations\cite{KOWALIK}
\begin{eqnarray}
\frac{dx}{dt} & = & rx(1-x^w)-(1-r)z \nonumber \\
\frac{dy}{dt} & = & (1-r)z(1-y^w)-(1-r)(y-z) \label{mfa2} \\
\frac{dz}{dt} & = &
\frac{rx(1-x^w)(y-z)}{1-x}-\frac{(1-r)z(1+z-x-y)(1-y^w)}{1-y}
-(1-r)zy^w \nonumber
\end{eqnarray}
In the first term of the third equation of Eqs.~(\ref{mfa2})
$x(1-x^w)$ gives the probability that the chosen site contains the
prey and at least one of its neighbours does not. The factor
$\frac{y-z}{1-x}$ gives the probability that the site chosen for
reproduction of the prey is occupied by the predator only. The set
of equations (\ref{mfa2}) remains in a much better (than
(\ref{mfa1})) agreement with Monte Carlo
simulations\cite{KOWALIKPHD}. In particular it predicts,
oscillatory regime for $w\geq 4$. In Monte Carlo simulations of
models on cartesian lattices (we made simulations only for
$d=1,2,3$) oscillations appear only in the $d=3$ case (i.e.,
$w=6$), and for $d=2$ (i.e., $w=4$) these were only
quasi-oscillations with the amplitude that vanishes in the limit
$N\rightarrow \infty$.

\begin{figure}
\vspace{2.5cm}
\centerline{\hspace{2cm}\psfig{file=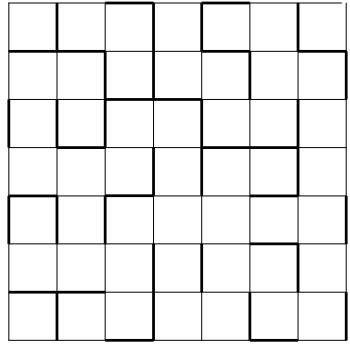,width=9cm,angle=0}}
\vspace{-4cm}\caption{In the bond percolation problem a fraction
$c$ of bonds is occupied (thick lines)\cite{STAUFFER1,STAUFFER2}.
Neighbouring occupied bonds form clusters. When $c$ exceeds a
certain threshold value an infinite (i.e., spanning the entire
lattice) cluster is formed. In a related problem of site
percolation a fraction of sites is occupied and neighbouring
occupied sites form clusters. As suggested by Rosenfeld \emph{et
al.}\cite{ROSENFELD,MONETTI} oscillatory behaviour might be
induced by some kind of percolation transition i.e., formation of
an infinite cluster of preys or predators. Although the idea is
appealing some calculations do not support it\cite{KOWALIKPHD}.}
\label{perkol}
\end{figure}

It is interesting to ask what is the mechanism that triggers the
emergence of finite-amplitude oscillations. Rosenfeld \emph{et
al.} suggested\cite{ROSENFELD,MONETTI} that oscillatory behaviour
in another lattice prey-predator model is induced by some kind of
percolation transition (see Fig.~\ref{perkol}). However, precise
measurements of cluster properties in our model has shown that
although some percolation transitions are indeed close to the
onset of oscillatory regime, they clearly do not overlap with this
onset\cite{KOWALIKPHD}. Another proposal relates oscillations with
some kind of stochastic resonance\cite{LIP1999, RAI}. Such a
relation might be suggested by the mean-field approximation
(\ref{mfa1}) that in fact describes a quasi-oscillatory dynamical
system. Stochastic fluctuations, that are present in the lattice
model but are neglected in the mean-field description, might be
considered as a noise perturbing such a dynamical system. As shown
by Gang et al.\cite{GANG} due to stochastic resonance, in some
low-dimensional autonomous dynamical systems noise might induce
oscillatory behaviour and one can expect that a similar scenario
operates in lattice prey-predator systems.

%%%%%%%%%%%%%%%%%%%%%%%%%%%%%%%%%%%%%%%%%%%%%%%%
%%%%%%%%%%%%%%%%%%%%%%%%%%%%%%%%%%%%%%%%%%%%%%%%

There are also other lattice prey-predator models were similar
oscillations were
observed\cite{MATSUDA,SATULOVSKY,PROVATA,ANTAL,BOCCARA,PEKALSKI}.
In a more general model, were predation and reproduction time
scales are independent, a first-order phase transition might
appear\cite{MOBILIA}. One can also mention that there are some
important ecological problem that so far were not examined with
individual based modelling but where such an approach might prove
to be valuable. In this context one can mention various
synchronization problems in spatially extended ecological
systems\cite{BLASIUS} and in particular the Moran effect
describing synchronization of populations exposed to common
noise\cite{ENGEN}.
\section{Modelling of complex ecosystems}\label{multi}
Models that we discussed in the previous section describe rather
simple ecosystems composed of few (two, three,\dots) species.
Dynamics of such models implements basic ecological processes:
reproduction, death, and in some cases also migration or aging. In
such models changes of the populations  takes place on a
characteristic ecological time scale that is set by the dynamics
of the models Typically, in real ecosystems this scale is of the
order of years (for example, in a hare-lynx system oscillations
with the period of approximately 10 years were
identified\cite{SCHAFFER}). But there are also some other than
ecological processes. On the so-called evolutionary (or
geological) time scale the entire species might die, change, or
give rise to a new species. The evolutionary time scale is usually
considered as much longer than ecological one\cite{GINGERICH}. As
a result very often researchers constructed specific models
directed toward either ecological or evolutionary processes.
However, there are numerous examples showing that these time
scales are not that much different and in some cases they are even
comparable\cite{THOMPSON,FUSSMANN}. Thus, such a separation of
time scales is to some extent artificial and was used mainly for
the ease of modelling (for a theoretical discussion of some
related issues see e.g., the paper by Khibnik and
Kondrashov\cite{KHIBNIK1997}). Actually, the complexity of real
multi-species ecosystems and the difficulty to model them to some
extent follow from the fact that these scales are not completely
separated and ecological and evolutionary processes affect each
other.

\begin{figure}
\vspace{3cm} \centerline{\hspace{3cm}
\psfig{file=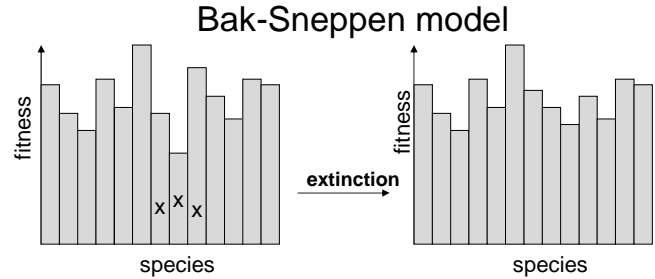,width=13cm,angle=0}}
\vspace{-8cm}\caption{The one-dimensional version of the
Bak-Sneppen model. First, the species of lowest fitness is
selected. Then this species and two of its neighbours (denoted by
crosses) go extinct and are replaced by three new species with
randomly selected fitness. Such a simple dynamics (no control
parameters) drives the model into the critical state. The extremal
dynamics of the Bak-Sneppen model was criticized on biological
grounds, nevertheless this model drew considerable attention and
is one of the main models of the self-organized criticality.}
\label{bak-sneppen}
\end{figure}

A model of multi-species ecosystem that tries to describe
evolutionary processes and drew considerable attention especially
in physicists community was introduced by Bak and
Sneppen\cite{BAKSNEPP1993} (see Fig.~\ref{bak-sneppen}). An
interesting property of this model is its self-organized
criticality. Namely, dynamics drives the model into the state
where extinctions are strongly correlated (like in critical
systems) and such a behaviour resembles the punctuated equilibrium
hypothesis of Eldredge and Gould\cite{ELDREDGE1972,SIMPSON1983}.
However, the Bak-Sneppen model has the dynamics that is operating
at the level of species and refers to the (still controversial)
notion of fitness. Thus, the model neglects ecological effects and
despite rich and intriguing dynamics can be considered only as a
toy model of an ecosystem. Nevertheless, the work of Bak and
Sneppen inspired other researchers to examine a number of models
with species-level dynamics. For example, Vandevalle and Ausloos
incorporated speciation\cite{VANDEVALLE}, Sol\'e and Manrubia
introduced various interactions between species\cite{SOLE}, and
Amaral and Meyer considered some elements of the food-chain
dynamics\cite{AMARAL}. Although these models drastically simplify
the dynamics of real ecosystems they do provide a valuable
qualitative description of some complex problems such as formation
of trophic levels or correlations and intensity of speciation and
extinction events\cite{NEWMANPALMER2003}.

%%%%%%%%%%%%%%%%%%%%%%%%%%%%%%%%%%%%%%%%%%%%%%%%
%%%%%%%%%%%%%%%%%%%%%%%%%%%%%%%%%%%%%%%%%%%%%%%%

\begin{figure}
\centerline{\psfig{file=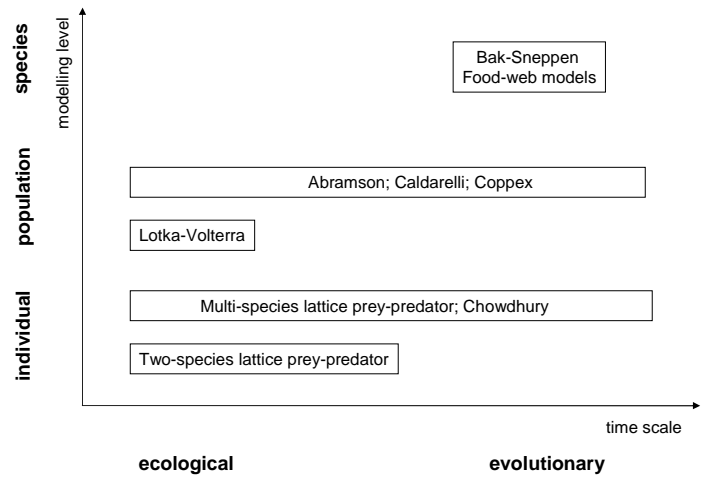,width=9cm,angle=270}}
\caption{Ecological and evolutionary aspects of the modelling of
an ecosystem. The most coarse-grained models of ecosystem have
dynamics operating at the level of species. Such models (e.g.,
Bak-Sneppen\cite{BAKSNEPP1993}, Food-web models\cite{AMARAL})
neglect population level processes and describe ecosystem at the
evolutionary time scale. Models originating from the
Lotka-Volterra model use dynamics defined at the population level.
Such models describe ecological processes in few-species
ecosystems, but multi-species versions of Abramson\cite{ABRAMSON},
Caldarelli et al.\cite{CALDARELLI} or Coppex et al.\cite{COPPEX},
that encompass also the evolutionary process were examined as
well. Similar range of applicability have models with individual
level dynamics (the model of Chowdhury et al.\cite{CHOWDHURY}
neglects, however, the heterogeneities in spatial distribution of
organisms).} \label{timescales1}
\end{figure}

Recently, computational methods made feasible the analysis of
models that incorporate both ecology and evolution. One way to
construct such models is to generalize Lotka-Volterra equations to
the multi-species case and to implement some speciation and
extinction mechanism. Such an approach has already been
developed\cite{ABRAMSON,COPPEX,CALDARELLI}, but it has similar
drawbacks as original Lotka-Volterra model, namely it neglects
spatial heterogeneities. In an alternative approach one uses
individual-based dynamics and some models of multi-species
ecosystems equipped with such a dynamics were
examined\cite{CHOWDHURY}. A diagram that illustrates some types of
models and their range of applicability is shown in
Fig.\ref{timescales1}.
%%%%%%%%%%%%%%%%%%%%%%%%%%%%%%%%%%%%%%%%%%%%%%%%%%%%%%%%%%%%
%%%%%%%%%%%%%%%%%%%%%%%%%%%%%%%%%%%%%%%%%%%%%%%%%%%%%%%%%%%%%
\section{Multispecies prey-predator model and periodicity of extinctions}\label{extin}
In this section we describe the multi-species version of a lattice
prey-predator model\cite{LIP2005,LIPLIP2006}. Numerical
simulations of the model show that the periodicity of mass
extinctions, that was suggested by Raup and
Sepkoski\cite{RAUPSEPKOSKI1984}, might be a natural feature of the
ecosystem's dynamics and not the result of a periodic external
perturbation.
\subsection{Model}\label{model2005}
Our model might be considered as a generalization of the
two-species model described in section~\ref{lattice}. The model is
defined on a $d$-dimensional cartesian lattice of the linear size
$N$. Similarly to the two-species model one uses the four-state
variables $\epsilon_i=0,1,2$ or 3. In addition, each predator is
characterized by its size $m$ ($0 < m < 1$) that determines its
consumption rate and at the same time its strength when it
competes with other predators. Only approximately the size $m$ can
be considered as related with physical size. Predators and preys
evolve according to rules typical to such systems (e.g., predators
must eat preys to survive, preys and predators can breed provided
that there is an empty site nearby, etc.). In addition, the
relative update rate for preys and predators is specified by the
parameter $r$ ($0 < r < 1$) and during breeding mutations are
taking place with the probability $p_{mut}$. More detailed
definition of the model dynamics is given below:

\begin{itemize}
\item Choose a site at random (the chosen site is denoted by $i$).
\item Provided that $i$ is occupied by a prey (i.e., if
$\epsilon_i = 1$ or $\epsilon_i = 3$) update the prey with the
probability $r$. If at least one neighbor (say $j$) of the chosen
site is not occupied by a prey (i.e., $\epsilon_j = 0$ or
$\epsilon_j = 2$), the prey at the site $i$ produces an offspring
and places it on an empty neighboring site (if there are more
empty sites, one of them is chosen randomly). Otherwise (i.e., if
there are no empty sites) the prey does not breed. \item Provided
that $i$ is occupied by a predator (i.e., $\epsilon_i = 2$ or
$\epsilon_i = 3$) update the predator with the probability
$(1-r)m_i$, where $m_i$ is the size of the predator at site $i$.
If the chosen site $i$ is occupied by a predator only ($\epsilon_i
= 2$), it dies, i.e., the site becomes empty ($\epsilon_i = 0$).
If there is also a prey there ($\epsilon_i = 3$), the predator
consumes the prey (i.e., $\epsilon_i$ is set to 2) and if
possible, it places an offspring at an empty neighboring site. For
a predator of the size $m_i$ it is possible to place an offspring
at the site $j$ provided that $j$ is not occupied by a predator
($\epsilon_j = 0$ or $\epsilon_j = 1$) or is occupied by a
predator ($\epsilon_j = 2$ or $\epsilon_j = 3$) but of a smaller
size than $m_i$ (in such a case the smaller-size predator is
replaced by an offspring of the larger-size predator). The
offspring inherits its parent's size with the probability
$1-p_{mut}$ and with the probability $p_{mut}$ it gets a new size
that is drawn from a uniform distribution.
\end{itemize}

At first sight one can think that such a model describes an
ecosystem with two trophic levels (preys and predators) and only
with predators being equipped with evolutionary abilities, which
would be of course highly unrealistic. Let us notice, however,
that expansion of predators sometimes proceeds at the expense of
smaller-size predators. Thus, predators themselves are involved in
prey-predator-like interactions. Perhaps it would be more
appropriate to consider unmutable preys as a renewable (at a
finite rate) source of, e.g., energy, and predators as actual
species involved in various prey-predator interactions and
equipped with evolutionary abilities.

In the remaining part of this section we will describe a possible
application of our model to the problem of mass extinctions and to
the problem of multiplicity of species in the Earth ecosystem as
contrasted with the uniqueness of the genetic code.
\subsection{Extinctions}
The suggestion that mass extinctions might be periodic in time was
made by Raup and Sepkoski\cite{RAUPSEPKOSKI1984}. While analyzing
fossil data, they noticed that during the last 250 My (million
years) mass extinctions on Earth appeared more or less cyclically
with a period of approximately 26My. Although their analysis was
initially questioned\cite{PATTERSON}, some other works confirmed
Raup and Sepkoski's hypothesis\cite{FOX,PROKOPH,PLOTNICK}. The
suggested large periodicity of mass extinctions turned out to be
very difficult to explain. Indeed, 26My does not seem to match any
of known Earth cycles and some researchers have been looking for
more exotic explanations involving astronomical
effects\cite{DAVIS,RAMPINO}, increased volcanic
activity\cite{STOTHERS1993}, or the Earth's magnetic field
reversal\cite{STOTHERS1986}. So far, however, none of these
proposals has been confirmed. One should also note that the most
recent analysis of palaeontological data that span last 542My
strongly supports the periodicity of mass extinctions albeit with
a larger cycle of about 62My\cite{ROHDE}.

Lacking a firm evidence of an exogenous cause, one can ask whether
the periodicity of extinctions might be explained without
referring to such a factor. In sections~\ref{ibm} and
\ref{lattice} we already mentioned that periodic behaviour of some
prey-predator systems is not the result of periodic driving but
rather a natural feature of their dynamics. However, the period of
oscillations in such systems is determined by the growth and death
rate coefficients of interacting species and is of the order of a
few years rather than tens of millions. Consequently, if the
periodicity of mass extinctions is to be explained within a model
of interacting species, a different mechanism that generates
long-period oscillations must be at work.

Such a mechanism might be at work in the multi-species model
described in section \ref{model2005}. Numerical
simulations\cite{LIP2005,LIPLIP2006} show that the model generates
long-period evolutionary oscillations The period of these
oscillations is determined by the inverse of the mutation rate and
we argued that it should be several orders of magnitude longer
than in the Lotka-Volterra oscillations. The mechanism that
generates oscillations in our model can be briefly described as
follows: A coevolution of predator species induced by the
competition for food and space causes a gradual increase of their
size. However, such an increase leads to the overpopulation of
large predators and a shortage of preys. It is then followed by a
depletion of large species and a subsequent return to the multi-
species stage with mainly small species that again gradually
increase their size and the cycle repeats. Numerical calculations
for our model show that the longevity of a species depends on the
evolutionary stage at which the species is created. A similar
pattern has been observed in some palaeontological
data\cite{MILLER} and, to our knowledge, the presented model is
the first one that reproduces such a dependence. Let us notice
that the oscillatory behaviour in a prey-predator system that was
also attributed to the coevolution has been already examined by
Dieckmann et al.\cite{DIECKMANN}. In their model, however, the
number of species is kept constant and it cannot be applied to
study extinctions. Moreover, the idea that an internal ecosystem
dynamics might be partially responsible for the long-term
periodicity in the fossil records was suggested by
Stanley\cite{STANLEY1990} and later examined by Plotnick and
McKinney\cite{PLOTNICK1993}. However, according to Stanley mass
extinctions are triggered by external impacts. Their approximately
equidistant separation is the result of a delayed recovery of the
ecosystem. In our approach no external factor is needed to trigger
such extinctions and sustain their approximate periodicity.

A gradual increase of size of species in our model recalls the
Cope's rule that states that species tend to increase body size
over geological time. This rule is not commonly accepted among
paleontologists and evolutionists and was questioned on various
grounds\cite{STANLEY1973}. However, recent studies of fossil
records of mammal species are consistent with this
rule\cite{ALROY,VANVALKEN}. Perhaps our model could suggests a way
to obtain a theoretical justification of this rule.

Although very complicated, in principle, it should be possible to
estimate the value of the mutation probability $p_{mut}$ from the
mutational properties of living species. Let us notice that in our
model mutations produce an individual that might be substantially
different from its parent. In Nature, this is typically the result
of many cumulative mutations and thus we expect that $p_{mut}$ is
indeed a very small quantity. Actually, $p_{mut}$ should be
considered rather as a parameter related with the speed of
morphological and speciation processes that are known to be
typically very slow\cite{GINGERICH}. Perhaps a different version
of the mutation mechanism where a new species would be only a
small modification of its parental species could be more suitable
for comparison with living species, but it might require longer
calculations.
\subsection{Unique genetic code and the emergence of a multi-species
ecosystem}\label{unique}
All living cells use the same code that is
 responsible for the transcription of information from DNA to
proteins\cite{ORGEL,SZATHMARY}. It suggests that at a certain
point of evolution of life on Earth a replicator that invented
this apparently effective mechanism was able to eliminate
replicators of all other species (if they existed) and establish,
at least for a short time, a single-species ecosystem. Although
this process is still to a large extent mysterious, one expects
that subsequent evolution of these successful replicators leads to
their differentiation and proliferation of species. In such a way
the ecosystem shifted from a single- to multi-species
one\cite{LIP2000}. It seems to us that our model might provide
some insight into this problem. Numerical simulations
show\cite{LIP2005,LIPLIP2006} that the oscillatory behaviour
appears in our model only for the relative update rate $r < 0.27$.
When preys reproduce faster ($r
> 0.27$), a different behaviour can be seen and
the model reaches a steady state with almost all predators
belonging to the same species with the size $m$ close to 1. Only
from time to time a new species is created with even larger $m$
and a change of the dominant species might take place. In our
opinion, it is possible that at the very early period of evolution
of life on Earth, the ecosystem resembled the case $r > 0.27$.
This is because at that time substrates ('preys') were renewable
faster than primitive replicators ('predators') could use them. If
so, every invention of the increase of the efficiency ('size')
could invade the entire system. In particular, the invention of
the coding mechanism could spread over the entire system. A
further evolution increased the efficiency of predators and that
effectively shifted the (single-species) ecosystem toward the $r <
0.27$ (multi-species, oscillatory) regime.
\subsection{Multispecies prey-predator model - summary and
perspectives} In this section we discussed a model where densities
of preys and predators as well as the number of species show
long-term oscillations, even though the dynamics of the model is
not exposed to any external periodic forcing. It suggests that the
oscillatory behaviour of the Earth ecosystem predicted by Raup and
Sepkoski could be simply a natural feature of its dynamics and not
the result of an external factor. Some predictions of our model
such as the lifetime of species or the time dependence of their
population sizes might be testable against palaeontological data.
Certainly, our model is based on some restrictive assumptions that
drastically simplify the complexity of the real ecosystem.
Nevertheless, it includes some of its important ingredients:
replication, mutation, and competition for resources (food and
space). As an outcome, the model shows that typically there is no
equilibrium-like solution and the ecosystem remains in an
evolutionary cycle. The model does not include geographical
barriers but let us notice that palaeontological data that suggest
the periodicity of mass extinctions are based only on marine
fossils\cite{ROHDE}. More realistic versions should take into
account additional trophic levels, gradual mutations, or sexual
reproduction. One should also notice that the palaeontological
data are mainly at a genus, and not species level. It would be
desirable to check whether the behaviour of our model is in some
sense generic or it is merely a consequence of its specific
assumptions. An interesting possibility in this respect could be
to recast our model in terms of Lotka-Volterra like equations and
use the methodology of adaptive dynamics developed by Dieckmann et
al.\cite{DIECKMANN}. Of course, the real ecosystem was and is
exposed to a number of external factors such impacts of
astronomical objects, volcanism or climate changes. Certainly,
they affect the dynamics of an ecosystem and contribute to the
stochasticity of fossil data. Filtering out these factors and
checking whether the main evolutionary rhythm is indeed set by the
ecosystem itself, as suggested in the present paper, is certainly
a difficult task but maybe worth an effort.
%%%%%%%%%%%%%%%%%%%%%%%%%%%%%%%%%%%%%%%%%%%%%%%%%%%%%%%%%%%%%%%%%%%%%%%
\section{Computational approaches to the evolution of language}\label{lang}
In this section we describe computational approaches to the
problem of evolution of language. In this field the mainstream
research takes the darwinian standpoint: natural selection guided
the language development and emergence of its basic features. One
thus accept that at least some features of language have certain
adaptive value and their gradual development is much more
plausible than a catastrophic change. There are, however, some
issues that still remain unclear within such a darwinian approach.
For example evolutionary  development of a reliable communication
system requires a substantial amount of altruism and it is not
clear whether standard explanations  that refer to kin selection
or reciprocal altruism are applicable (for example kin selection
does not explain our willingness to talk to non-kin). Another
problem is concerned with the interaction of evolution and
learning, sometimes known as a Baldwin effect. In some cases,
learning is known to direct the evolutionary changes, and perhaps
in such a way humans developed a language-specific adaptations
commonly termed Language Acquisition Device (LAD). Efficiency of
the Baldwin effect and even the very existence of LAD remain,
however, open problems. There is perhaps a little chance that
computational modelling will definitely resolve these issues. But
already at the level of constructing appropriate models one has to
quantify relevant processes and effects, and even that can provide
a valuable insight.
\subsection{Evolution and language development}\label{lang-evol}
The ability to use language distinguishes humans from all other
species. Certain species also developed some communication modes
but of much smaller capabilities as well as complexity.  Since
several decades various schools are trying to explain the
emergence and development of language. Nativists argue that
langauge capacity is a collection of domain-specific cognitive
skills that are somehow encoded in our genome. However, the idea
of the existence  of such a Language Acquisition Device or
"language organ" (the term coined by their most prominent
representative Noam Chomsky\cite{CHOMSKY}), was challenged by
empiricists, who argue that linguistic performance of humans can
be explained using domain-general learning techniques. The recent
critique along this line was made by Sampson\cite{SAMPSON}, who
questions even the most appealing argument of nativists, that
refer to the poverty of stimulus and apparently fast learning of
grammar by children. An important issue of possible adaptative
merits of language does not seem to be settled either.
Non-adaptationists, again with Chomsky as the most famous
representative\cite{CHOMSKY1972}, consider language as a side
effect of other skills and thus claim that its evolution, at least
at the beginning, was not related with any fitness advantage. A
chief argument against the non-adaptationist stand is the
observation that there is a number of costly adaptations that seem
to support human linguistic abilities such as a large brain, a
longer infancy period or descended larynx. Recently, in their
influential paper, Pinker and Bloom argued that, similarly to
other complex adaptations, language evolution can only be
explained by means of natural selection
mechanisms\cite{PINKER1990}. Their paper triggered a number of
works where language was examined from the perspective of
evolutionary biology or game theory\cite{JACKENDOFF,KNIGHT}. In
particular, Nowak et al. used some optimization arguments, that
might explain the origin of some linguistic
universals\cite{NOWAK,NOWAK1999}. They suggest that words appeared
in order to increase the expressive capacity and sentences (made
of words) limit memory requirements. Confrontation of nativists
with empiricists and adaptationists with non-adaptationists so far
does not seem to lead to consensus but certainly deepened our
understanding of these problems\cite{SMITH}.

Recently, a lot of works on the language emergence seem to have an
evolutionary flavour. Such an approach puts some constraints on
possible theories of the language origin. In particular, it rules
out non-adaptationist theories, where language is a mere
by-product of having a large and complex brain\cite{GOULD}. The
emergence of language has been also listed as one of the major
transitions in the evolution of life on Earth\cite{MSMITH1997}. An
interesting question is whether this transition  was variation or
selection limited\cite{SZAMADO}. In variation limited transitions
the required configuration of genes is highly unlikely and it
takes a considerable amount of time for the nature to invent it.
For selection limited transitions the required configuration is
easy to invent but there is no (or only very weak) evolutionary
pressure that would favour it. Relatively large cognitive
capacities of primates and their genetic proximity with humans
suggests that some other species could have been also capable to
develop language-like communication. Since they did not, it was
perhaps due to a weak selective pressure. Such indirect arguments
suggest that the emergence of language was selection
limited\cite{SZAMADO}.

Some interesting results can be obtained by applying
game-theory\index{game theory} reasoning to one of the most basic
problems of emerging linguistic communication, namely why do we
talk (at all!) and why do we exchange valuable and trustful
information. Since speaking is costly (it takes time, energy and
sometimes might expose a speaker to predators), and listening is
not, such a situation seems to favour selfish individuals that
would only listen but would not speak. Moreover, in the case of
the conflict of interests the emerging communication system would
be prone to misinformation or lying. The resolution of these
dilemmas usually refers to the kin selection\cite{HAMILTON} or
reciprocal altruism\cite{TRIVERS}. In other words, speakers remain
honest because they are helping their relatives or they expect
that others will do the same for them in the future. As an
alternative explanation Dessalles\cite{DESSALLES} suggests that
honest information is given freely because it is profitable - it
is a way of competing for status within a group. Computational
modelling of Hurford\cite{HURFORD} gives further evidence that
speaking might be more profitable than listening. Hurford
considered agents engaged in communicative tasks (one speaker and
one hearer) and their abilities evolved with the genetic algorithm
that was set to prefer either communicative or interpretative
success. Only in the former case the emerging language was similar
to natural languages were synonymy was rare and homonymy
tolerated. When interpretative success was used as the basis of
selection then the converse situation (unknown in natural
language) arose: homonymy was rare and synonymy tolerated. Some
related results on computational modelling of the honest cost-free
communication are reported by Noble\cite{NOBLE}.

A necessary ingredient of language communication is
learning\index{learning}. It is thus legitimate to ask whether
darwinian selection might be responsible for the genetic
hard-wiring of a Language Acquisition Device. Indeed, this (to
some extent hypothetical) organ is most likely responsible for
some of the arbitrary (as opposed to the functional) linguistic
structures.  But for such an organ to be of any value, an
individual first has to acquire the language. The inheritance of
characteristics acquired during an individual lifetime is usually
associated with discredited lamarckian mechanism and thus
considered to be suspicious. However, the relation between
evolution and learning is more delicate and the attempts to
clarify the mutual interactions of these two adaptive mechanisms
have a long history. According to a purely darwinian explanation,
known as a Baldwin
effect\cite{BALDWIN1896,SIMPSON,WEBER}\index{Baldwin effect},
there might appear a selective pressure in a population for the
evolution of the instinctive behaviour that would replace the
beneficial, but costly, learned behaviour\cite{TURNEY1996}.
Baldwin effect presumably played an important role in the
emergence and evolution of language but certain aspects of these
processes still remain unresolved\cite{YAMA2004}. For example, one
of the assumptions that is needed for the Baldwin effect to be
effective is a relatively stable environment since otherwise
rather slow evolutionary processes will not catch up with the fast
changing environment. Since the language formation processes are
rather fast (in comparison to the evolutionary time scale),
Christiansen and Chater questioned the role of adaptive
evolutionary processes in the formation of arbitrary structures
like Language Acquisition Device\cite{CHRIST}. Actually, they
suggest a much different scenario, where it is a language that
adapted to human brain structures rather than vice versa.
\subsection{Language as a complex adapting system}\label{lang-adap}
\begin{figure}%[ht]
\vspace{4.0cm} \centerline{ \epsfxsize=13cm \epsfbox{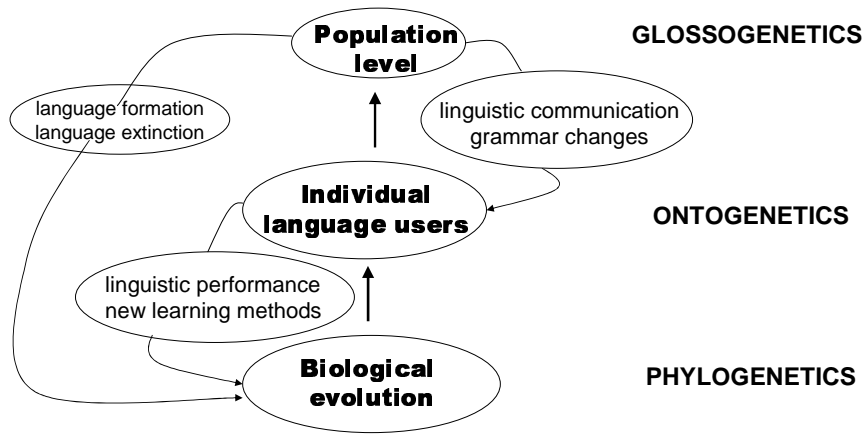} }
\vspace{-4cm} \caption{Language as a complex adaptive system.Many
different processes governing the language evolution are entangled
at various levels. Relatively fast individual level
(ontogenetics), comprising e.g., langauge acquisition processes,
is determined mainly by interactions between individual language
users. Much slower are populational-level processes
(glossogenetics) such as language formations, extinctions, grammar
changes or migrations. To obtain a complete description one has to
consider also biological evolution (phylogenetics) and these are
the slowest processes of the language evolution. Various processes
at individual and population level affect the fitness landscape
and that influences the biological evolution level. Similarly,
individual language user level is affected by populational level
processes.} \label{fig1}
\end{figure}

From the above description it is clear that studying of the
emergence and evolution of language is a complex and
multidisciplinary task  and requires cooperation of not only
linguists, neuroscientists, and anthropologists, but also  experts
in artificial intelligence, computer sciences or evolutionary
biology\cite{NOWAK2002}. One can distinguish two levels at which
language can be studied and described\cite{DEBOER}
(Fig.~\ref{fig1}). At the individual level the description centers
on the individual language users: their linguistic performance,
language acquisition, speech errors, speech pathologies or brain
functioning in relation with language processing. At the
individual level the language of each individual is slightly
different. Nevertheless, within certain population these
individuals can efficiently communicate and that establishes the
population level. At this level the language is considered as an
abstract system that exists in a sense separately from the
individuals users. There are numerous interactions between these
two levels. Indeed, the linguistic behaviour of individuals
depends on the language (at the population level) specific to the
population they are part of. And, as a feedback, the language used
in a given population is a collective behaviour and emerges from
linguistic behaviour of individuals composing this population.
Various processes shaping such a complex system are operating at
different time scales. The fastest dynamics is operating at the
individual level (ontogenetic timescale\cite{KIRBY2002}) that
includes, for example, language acquisition processes. Much slower
processes, such as migrations of language populations, dialects
formation or language extinctions, are operating at the so-called
glossogenetic timescale. The slowest processes govern the
biological evolution of language users and that defines the
phylogenetic timescale. Processes operating at these different
timescales are not independent (Fig.~\ref{fig1}). Biological
evolution might change linguistic performance of individuals and
that might affect the glossogenetic processes. For example, a
mutation that changes the vocal ability of a certain individual,
if spread in his/her population, might lead to a dialect formation
or a language extinction. Such population-level processes might
change the selective pressure that individual language users are
exposed to and that might affect phylogenetic processes, closing
thus the interaction loop.

Various levels of descriptions and processes operating at several
timescales suggest that complex models must be used to describe
adequately the language evolution. Correspondingly, the analysis
of such models and predicting their behaviour also seem to be
difficult. It is known that some phenomena containing feedback
interactions might be described in terms  of nonlinear
differential equations, such as for example already described
Lotka-Volterra equations. The behaviour of such nonlinear
equations is often difficult to predict, since abrupt changes even
of the qualitative nature of solutions might take place. Language
evolution is, however, much more complex than ecological problems
of interacting populations and its description in terms of
differential equations would be much more complicated if at all
feasible. %%%%%%%%%%%%%%%%%%%%%%%%%%%%%%%%%%%%%%%%%%%%%%%%%%%%%%
%%%%%%%%%%%%%%%%%%%%%%%%%%%%%%%%%%%%%%%%%%%%%%%%%%%%%%%%%%%%%%%%%
It seems that recently the most promising and frequently used
approach to examine such systems is  computational modelling of
multi-agent systems\index{multi-agent systems}. Using this method
one examines a language that emerges in a bottom-up fashion as a
result of interactions within a group of agents equipped with some
linguistic functions. Then one considers language as a complex
adaptive system that evolves and complexifies according to
biologically inspired principles such as selection and
self-organization\cite{STEELS2002}. Thus, the emerging language is
not static but evolves in a way that hopefully is similar to human
language evolution. Of course, using such an approach one cannot
explain all intricacies of human languages. A more modest goal
would be to understand some rather basic features that are common
to all languages such as meaning-form mappings, origin of
linguistic coherence (among agents without central control and
global view), or coevolutionary origin of grammar and meaning.

Within such a multi-agent approach, two groups of models can be
distinguished. In the first one, originating from the so-called
iterated learning model\index{iterated learning model}, one is
mainly concerned with the transmission of language between
successive generations of agents\cite{KIRBY2001,BRIGHTON2002}.
Agents that are classified as teachers produce some expressions
that are passed to learners that try to infer their meaning using
statistical learning techniques such as neural networks. After a
certain number of iterations teachers are replaced by learners and
a new population of learners is introduced. The important issue
that the iterated learning model has successfully addressed is the
transition from holistic (complex meaning expressed by a single
form) to compositional language (composite meaning is expressed
with composite form). However, since such a procedure is
computationally relatively demanding and the number of
communicating agents is thus typically very small, the problem of
the emergence of linguistic coherence must be neglected in this
approach. To tackle this problem Steels introduced a naming game
model\cite{STEELS1995}\index{naming game}. In this approach one
examines a population of agents trying to establish a common
vocabulary for a certain number of objects present in their
environment. The change of generations is not required in the
naming game model since the emergence of a common vocabulary is a
consequence of the communication processes between agents, and
agents are not divided into teachers and learners but take these
roles in turn.
\subsection{Evolutionary naming game}\label{evol-name}
It seems that the iterated learning model and the naming-game
model are at two extremes: the first one emphasizes the
generational turnover while the latter concentrates on the
single-generation (cultural) interactions. Since in the language
evolution both aspects are present, it is desirable to examine
models that combine evolutionary and cultural processes. Recently
we have introduced such a model\cite{LIPLIP2008,LIPLIP2008A} and
below we briefly describe it properties.

In our model we consider a set of agents located at sites of the
square lattice of the linear size $N$. Agents are trying to
establish a common vocabulary on a single object present in their
environment. An assumption that agents communicate only on a
single object does not seem to restrict the generality of our
considerations and has already been used in some other studies of
naming game~\cite{BARONCHELLI2006,ASTA2006} or
language-change~\cite{NETTLE1999,NETTLE1999A} models. A randomly
selected agent takes the role of a speaker that communicates a
word chosen from its inventory to a hearer that is randomly
selected among nearest neighbours of the speaker. The hearer tries
to recognize the  communicated word, namely it checks whether it
has the word in its inventory. A positive or negative result
translates into communicative success or failure, respectively. In
some versions of the naming game
model~\cite{BARONCHELLI2006,ASTA2006} a success means that both
agents retain in their inventories only the chosen word, while in
the case of failure the hearer adds the communicated word to its
inventory.

To implement the learning ability we have modified this rule and
assigned weights $w_i$ ($w_i>0$) to each $i$-th word in the
inventory. The speaker selects then the $i$-th word  with the
probability $w_i/\sum_j w_j$ where summation is over all words in
its inventory (if its inventory is empty, it creates a word
randomly). If the hearer has the word in its inventory, it is
recognized. In addition, each agent $k$ is characterized by its
learning ability $l_k$ ($0<l_k<1$), that is used to modify
weights. Namely, in the case of success both speaker and hearer
increase the weights of the communicated word by their learning
abilities, respectively. In the case of failure the speaker
subtracts its learning ability from the weight of the communicated
word. If after such a subtraction a weight becomes negative, the
corresponding word is removed from the repository. The hearer in
the case of failure, i.e., when it does not have the word in its
inventory, adds the communicated word to its inventory with a unit
weight.

Apart from communication, agents in our model evolve according to
the population dynamics: they can reproduce, mutate, and
eventually die. To specify intensity of these processes we have
introduced the communication probability $p$. With the probability
$p$ the chosen agent becomes a speaker and with the probability
$1-p$ a population update is attempted. During such a move the
agent dies with the probability $1-p_{\rm surv}$, where $p_{\rm
surv}=\exp (-at)[ 1-\exp(-b\sum_j w_j/\langle w\rangle)]$, and
$a\sim 0.05$ and $b=5$ are certain parameters whose role is to
ensure a certain speed of population turnover. Moreover, $t$ is
the age of an agent and $\langle w\rangle$ is the average (over
agents) sum of weights. Such a formula takes into account both its
linguistic performance (the bigger $\sum_j w_j$ the larger $p_{\rm
surv}$) and its age. If the agent survives (it happens with the
probability $p_{\rm surv}$), it breeds, provided that there is an
empty site among its neighbouring sites. The offspring typically
inherits parent's learning ability and the word from its inventory
that has the highest weight. In the offspring's inventory the
weight assigned initially  to this word equals one. With the small
probability $p_{\rm mut}$ a mutation takes place and the learning
ability of an offspring is selected randomly anew. With the same
probability an independent check  is  made whether to mutate the
inherited word. Numerical simulations show that the described
below behaviour of our model is to some extent robust with respect
to some modifications of its rules. For example, qualitatively the
same behaviour is observed for modified parameters $a$ and $b$,
different form of the survival probability $p_{\rm surv}$
(provided it is a decreasing function of $t$ and an increasing
function of $\sum_j w_j$), or different breeding and/or
mutation\index{mutation} rules. To examine the behaviour of the
model we have measured the communication success rate $s$ defined
as an average over agents and simulation time of the fraction of
successes with respect to all communication attempts. Moreover, we
have measured the average learning ability $l$.

%%%%%%%%%%%%%%%%%%%%%%%%%%%%%%%%%%%%%%%%%%%%%%%%%%%%%%%%%%
%\begin{figure}
\begin{figure}
\vspace{2.5cm} \centerline{ \epsfxsize=13cm \epsfbox{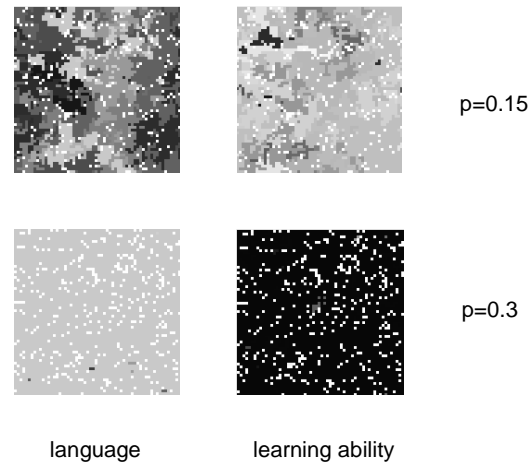}
}
%\figspace
\vspace{-5cm} \caption{Exemplary configurations of the
evolutionary naming game model\cite{LIPLIP2008,LIPLIP2008A} with
$L=60$ and $p_{\rm mut}=0.001$. In the small-$p$ phase (upper
panel) communications are infrequent and agents using the same
language (left)or having the same learning abilities (right) form
only small clusters. In this phase the communication success rate
$s$ and the learning ability $l$ are small (see also
Fig.~\ref{zbaszyn}). The larger the learning ability of an agent
the darker are pixels representing it (white: l=0; black: l=1). In
the large-$p$ phase (lower panel) frequent communications result
in the emergence of the common language. Moreover, almost all
agents use the same language and have the same, and large,
learning ability.}\label{conf-new}
\end{figure}
%%%%%%%%%%%%%%%%%%%%%%%%%%%%%%%%%%%%%%%%%%%%%%%%%%%%%%%%%%%%%%%%%%%%%%%%%%%%%%
%%%%%%%%%%%%%%%%%%%%%%%%%%%%%%%%%%%%%%%%%%%%%%%%%%%%%%%%%%
%\begin{figure}
\begin{figure}
\vspace{1cm} \centerline{ \epsfxsize=9cm \epsfbox{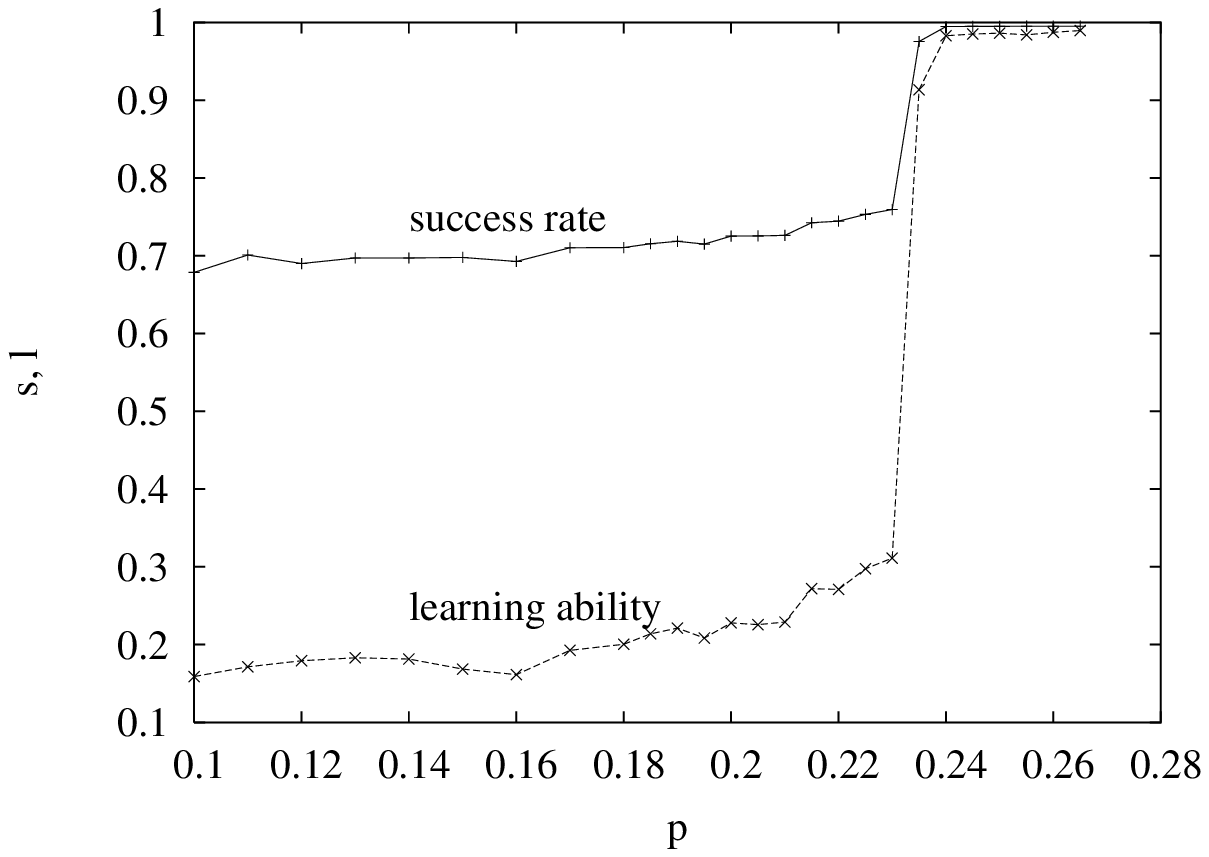} }
%\figspace
\vspace{5mm} \caption{The success rate $s$ and the learning
ability $l$ as a function of the communication probability $p$.
Calculations were made for system size $L=60$ and mutation
probability $p_{\rm mut}=0.001$. Simulation time for each value of
$p$ was typically equal to $10^5$ steps with $3\cdot 10^4$ steps
discarded for relaxation. A step is defined as a single, on
average, update of each site.} \label{zbaszyn}
\end{figure}
%%%%%%%%%%%%%%%%%%%%%%%%%%%%%%%%%%%%%%%%%%%%%%%%%%%%%%%%%%%%%%%%%%%%%%%%%%%%%%

Our model captures all three basic aspects of language: learning,
culture, and evolution. Agents in our model are equipped with an
evolutionary trait: learning ability. When communication between
agents is sufficiently frequent (i.e., when $p$ is large enough),
cultural processes create a niche in which a larger learning
ability becomes advantageous. It causes an increase of learning
ability, but its large value in turn makes the cultural processes
more efficient. As a result the model was shown to undergo an
abrupt bio-linguistic transition\index{bio-linguistic transition}
where both linguistic performance ($s$) and ability ($l$) of
agents change very rapidly (see
Figs.~\ref{conf-new}-\ref{zbaszyn})\cite{LIPLIP2008}. It was also
shown that under the plausible assumption, that the intensity of
communication increases continuously in time, this bio-linguistic
transition is replaced with a series of fast, transition-like
changes\cite{LIPLIP2008A}. In our opinion, the proposed model
shows that linguistic and biological processes have a strong
influence on each other, which has certainly contributed to an
explosive development of our species.
\subsection{Baldwin effect}\label{baldwin}
That learning in our model modifies the fitness landscape of a
given agent and facilitates the genetic accommodation of learning
ability is actually a manifestation of the much debated Baldwin
effect. The fact that the success rate $s$ and the learning
ability $l$ have a jump at the same value of $p$  shows that
communicative and biological ingredients in our model strongly
influence each other and that leads to the single and abrupt
transition. In our model successful communication requires
learning. A new-born agent communicating with some mature agents
who already  worked out a certain (common in this group) language
will increase the weight of a corresponding word. As a result, in
its future communications the agent will use mainly this word. In
what way such a learning might get coupled with evolutionary
traits? The explanation of this phenomenon is known as a Baldwin
effect. Although at first sight it looks like a discredited
Lamarckian phenomenon, the Baldwin effect is actually purely
Darwinian~\cite{HINTON1987,YAMA2004}. There are usually some
benefits related with the task a given species has to learn and
there is a cost of learning this task. One can argue that in such
case there is some kind of an evolutionary pressure that favours
individuals for which the benefit is larger or the cost is
smaller. Then, the evolution will lead to the formation of species
where the learned behaviour becomes an innate ability. It should
be emphasized that the acquired characteristics are not inherited.
What is inherited is the ability to acquire the characteristics
(the ability to learn)~\cite{TURNEY1996}. In the context of the
language evolution the importance of the Baldwin effect was
suggested by Pinker and Bloom~\cite{PINKER1990}. Perhaps this
effect is also at least partially responsible for the formation of
the Language Acquisition Device. However, many details concerning
the role of the Baldwin effect in the evolution of language remain
unclear~\cite{MUNROE2002}.

We already argued~\cite{LIPLIP2008},  that in our model the
Baldwin effect is also at work. Let us consider a population of
agents with the communication probability $p$ below the threshold
value ($p=p_c\approx 0.23$). In  such a case the learning ability
remains at a rather low level (since clusters of agents using the
same language are small, it does not pay off to be good at
learning the language of your neighbours). Now, let us increase
the value of $p$ above the threshold value. More frequent
communication changes the behaviour dramatically. Apparently,
clusters of agents using the same language are now sufficiently
large and it pays off to have a large learning ability because
that  increases the success rate and thus the survival probability
$p_{\rm surv}$. Let us notice that $p_{\rm surv}$ of an agent
depends on its linguistic performance ($\sum_j w_j$) rather than
its learning ability. Thus clusters of agents of good linguistic
performance (learned behaviour) can be considered as niches that
direct the evolution by favouring agents with large learning
abilities, which is precisely the Baldwin effect. It should be
noticed that linguistic interactions between agents (whose rate is
set by the probability $p$) are typically much faster than
evolutionary changes (set by $p_{\rm mut}$) and such an effect was
also observed in simulations~\cite{LIPLIP2008}.

As a result of a positive feedback (large learning ability
enhances communication that enlarges clusters  that favours even
more the increased learning ability) a discontinuous transition
takes place both with respect to the success rate and learning
ability . An interesting question is whether such a behaviour is
of any relevance in the context of human evolution. It is obvious
that development of language, which probably took place somewhere
around $10^5$ years ago, was accompanied by important anatomical
changes such as fixation of the so-called speech gene (FOXP2),
descended larynx or enlargement of brain~\cite{HOLDEN2004}.
Linguistic and other cultural interactions that were already
emerging in early hominid populations were certainly shaping the
fitness landscape and that could direct the evolution of our
ancestors via the Baldwin effect.

The examined model is not very demanding computationally. It seems
to be possible to consider agents talking on more than one
object\cite{LIPLIP2008B}, or to examine statistical properties of
simulated languages such as for example, distributions of their
lifetimes or of the number of users. It would be interesting to
examine the role of topology of interaction network and place
agents on complex networks, like e.g., scale-free networks, that
are known to provide a more realistic description of human
linguistic interactions\cite{ASTA2006}. One can also study
diffusion of languages, the role of geographical
barriers\cite{PATRIARCA}, or formation of language families. There
is already an extensive literature documenting linguistic data as
well as various computational approaches modelling, for example,
competition between already existing natural languages
~\cite{ABRAMS2003,STAUFFER2008,PAULO2007}. The dynamics of the
present model, that is based on an act of elementary
communication, offers perhaps more natural description of dynamics
of languages than some other approaches that often use some kind
of coarse-grained dynamics.
\section{Conclusions}\label{concl}
In the present paper we reviewed computational methods that are
used for modelling evolutionary systems. We emphasized the need
and advantages of using models with individual-based dynamics. We
also drew attention to various time scales of processes that shape
the evolution of complex systems. In ecosystems these are
ecological and evolutionary time scales. In the language evolution
cultural processes set an additional timescale. Perhaps the most
interesting phenomena arise from interactions of  processes of
various times scales. Evolutionary cycling or Baldwin effect are
excellent examples of such phenomena, to claim however their
satisfactory understanding, much remains to be done.

This mini-review is of course biased by our own experience in this
field. We did not even mention about a number of other approaches
and techniques of modelling evolutionary systems. Some of them are
covered in other chapters of this volume.
\section*{Acknowledgements}
The present paper is based on the lecture that A.L. delivered
during the conference "From Genetics to Mathematics"
(Zb\c{a}szy\'n, Poland, October-2007). A.L. thanks the organizers
of the conference  for invitation.  Our project is supported with
the research money allocated for the period 2008-2010 under the
grant N N202 071435. We also acknowledge access to the computing
facilities at Pozna\'n Supercomputer and Networking Center.

%\printindex[aindx]                 % to print author index
\printindex                         % to print subject index

\begin{thebibliography}{9}

\bibitem{ABRAMS2003} Abrams, D. and Strogatz, S.~H., (2003). Modelling the dynamics
of language death. \emph{Nature} {\bf 424}, p. 900.

\bibitem{ABRAMSON} Abramson, G. (1997). Ecological model of extinctions.
\emph{Phys. Rev. E} \textbf{55}, pp. 785-788.

\bibitem{ALROY} Alroy, J., (1998). Cope's Rule and the Dynamics of Body Mass
Evolution in North American Fossil Mammals. \emph{Science}
\textbf{280}, pp.~731--734.

\bibitem{AMARAL} Amaral, L.~A.~N. and Meyer, M. (1999).
Environmental changes, coextinction, and patterns in the fossil
record, \emph{Phys.~Rev.~Lett.} \textbf{82}, pp. 652-655.

\bibitem{ANTAL} Antal, T. and Droz, M. (2001).
Phase transitions and oscillations in a lattice prey-predator
model. \emph{Phys.~Rev.~E} \textbf{63} p.056119.

\bibitem{BAKSNEPP1993} Bak, P. and Sneppen, K. (1993), Punctuated
equilibrium and criticality in a simple model of evolution,
\emph{Phys.~Rev.~Lett.~}{\bf 71}, pp. 4083--4086.

\bibitem{BALDWIN1896} Baldwin, J.~M.~(1896). A new factor in
evolution. \emph{American Naturalist} \textbf{30}, pp. 441-451.

\bibitem{BARONCHELLI2006} Baronchelli, A., Felici, M., Loreto, V.,
Caglioti, E., and Steels, L., (2006). Sharp transition towards
shared vocabularies in multi-agent systems.
\emph{J.~Stat.~Mech.}~{\bf P06014}.

\bibitem{BLASIUS} Blasius, B., Huppert, A., and Stone, L. (1999).
Complex dynamics and phase synchronization in spatially extended
ecological systems. \emph{Nature} \textbf{399} pp. 354--359.

\bibitem{BOCCARA} Boccara, N., Roblin, O., and Roger, M.~(1994).
Automata network predator-prey model with pursuit and evasion.
\emph{Phys.~Rev.~E} \textbf{50}, pp. 4531--4541.

\bibitem{BRIGHTON2002} Brighton, H., (2002). Compositional Syntax from Cultural Transmission.
\emph{Artif.~Life} {\bf 8}, pp. 25--54.

\bibitem{CALDARELLI} Caldarelli, G., Higgs, P. G., and McKane, A. J. (1998). Modelling
coevolution in multispecies communities. \emph{J. Theor. Biol.}
\textbf{193}, pp. 345-358.

\bibitem{CHOMSKY} Chomsky, N., (1965). \emph{Aspects of the theory of syntax},
(Cambridge, MA: MIT Press).

\bibitem{CHOMSKY1972} Chomsky, N., (1972). \emph{Language and Mind},
(San Diego, Harcourt Brace Jovanovich).

\bibitem{CHOWDHURY} Chowdhury, D., Stauffer, D., and Kunwar, A. (2003). Uniffication of
Small and Large Time Scales for Biological Evolution: Deviations
from Power Law. \emph{Phys. Rev. Lett.} \textbf{90}, p.068101.

\bibitem{CHRIST} Christiansen M.~H. and Chater, N., (2007).
{\it Language as shaped by the brain}, preprint.

\bibitem{CHRISTIANSEN} Christiansen, M.~H.~and Kirby, S.~(2003).
Language Evolution: The Hardest Problem in Science. In
Christiansen, M.~H.~and Kirby, S. (eds.) \emph{Language Evolution}
(Oxford University Press Inc., New York).

\bibitem{COPPEX} Coppex, F., Droz, M., and Lipowski, A., (2004).
Extinction dynamics of Lotka-Volterra ecosystems on evolving
networks, \emph{Phys. Rev. E} \textbf{69}, p. 061901.

\bibitem{ASTA2006} Dall'Asta, L., Baronchelli, A., Barrat, A., and Loreto, V., (2006).
Nonequilibrium dynamics of language games on complex networks.
\emph{Phys.~Rev.~E} {\bf 74}, pp. 036105.

\bibitem{DAVIS} Davis, M., Hut, P., and  Muller, R.~M., (1984). Extinction of species
by periodic comet showers. \emph{Nature} \textbf{308}, pp.
715--717.

\bibitem{DAWKINS1976} Dawkins, R.~(1976). \emph{The
 Selfish Gene}, (Oxford University Press).

\bibitem{DEBOER} de Boer, B., (2006).
{\it Computer modelling as a tool for understanding language
evolution}. In {\it Evolutionary Epistemology, Language and
Culture - A nonadaptationist system theoretical approach},
Gonthier et al. (eds.) (Dordrecht: Springer).

\bibitem{DESSALLES} Dessalles, J.~L., (1998).
In J.~R.~Hurford, J.~R.~et al. (eds.), \emph{Approaches to the
Evolution of Language: Social and Cognitive Bases} (Cambridge
University Press, Cambridge).

\bibitem{DIECKMANN} Dieckmann, U., Marrow, P., and Law, R., (1995). Evolutionary Cycling
in Predator-Prey Interactions: Population Dynamics and the Red
Queen. \emph{J. Theor. Biol.} \textbf{176}, pp.~91--102.

\bibitem{ELDREDGE1972} Eldredge, N. and Gould, S.~J. (1972).
Punctuated Equilibria: An Alternative to Phyletic Gradualism. In
Schopf, T.~J.~M. (eds). \emph{Models in Palaeobiology} (Freeman,
Cooper, San Francisco) pp. 82-115.

\bibitem{ENGEN} Engen, S. and S\ae ther, B.-E. (2005). Generalizations of the Moran Effect
Explaining Spatial Synchrony in Population Fluctuations.
\emph{Amer.~Natur.}~\textbf{166}, pp. 603--612,

\bibitem{FOX} Fox, W.~T., (1987). Harmonic analysis of periodic
extinctions. \emph{Paleobiology} \textbf{13}, pp. 257--271.

\bibitem{FUSSMANN}Fussmann, G.~F., Ellner, S.~P., and Hairston Jr.,
N.~G. (2003). Evolution as a critical component of plankton
dynamics. \emph{Proc.~Roy.~Soc.~B} \textbf{270}, pp. 1015--1022.

\bibitem{GANG} Gang, H., Ditzinger, T., Ning, C.~Z., and Haken, H.~(1993). Stochastic Resonance without external periodic Force.
 \emph{Phys. Rev.~Lett.} \textbf{71}, pp. 807--810.

\bibitem{GINGERICH} Gingerich, P.~D., (1983). Rates of evolution: effects of time and
temporal scaling. \emph{Science} \textbf{222}, pp. 159–-161.

\bibitem{GOULD} Gould, S.~J. and Lewontin, R.~C. (1979). The Spandrels of San Marco and the
Panglossian Paradigm: A Critique of the Adaptationist Programme.
\emph{Proc.~Roy.~Soc.~London B} \textbf{205}, pp 581-598.

\bibitem{GRASSBERGER} Grassberger, P. and de la Torre, A. (1979). Reggeon field theory (Schl\"ogl's first model)
on a lattice; Monte Carlo calculations of critical behaviour,
\emph{Ann. Phys. (N.Y.)} \textbf{122}, pp. 373–-396.

\bibitem{GRINSTEIN} Grinstein, G., Mukamel, D., Seidin, R., and Bennett, Ch.~H. (1993).
Temporally periodic phases and kinetic roughening. \emph{Phys.
Rev. Lett.} \textbf{70}, pp. 3607--3610.

\bibitem{HAMILTON} Hamilton, W.~D., (1964). The genetical theory of social behaviour (I and II).
\emph{J.~Theor.~Biol.}~{\bf 7}, pp. 1--16.

\bibitem{HARDIN} Hardin, D.~P., Takac, P., and Webb, G.~F.~(1990).
Dispersion population models discrete in time and continuous in
space. \emph{J.~Math.~Biol.}~\textbf{28} pp. 1--20.

\bibitem{HASSELL} Hassell, M.~P., Comins, H.~N., and May, R.~M.
(1991). Spatial structure and chaos in insect population dynamics.
\emph{Nature, Lond.}~\textbf{353}, pp. 255-258.

\bibitem{HINRICHSEN} ~Hinrichsen, H., (2000). Nonequilibrium Critical Phenomena and Phase Transitions into
Absorbing States. \emph{Adv.Phys.} \textbf{49} pp. 815-958

\bibitem{HINTON1987} Hinton, G. and Nowlan, S. (1987). How learning can guide evolution.
\emph{Complex Systems} {\bf 1}, 495-502.

\bibitem{HOLDEN2004} Holden, C., (2004). The origin of Speech. \emph{Science} {\bf 303},
pp. 1316-1319.

\bibitem{HOLMES} Holmes, E.~E., Lewis, M.~A., Banks, J., and Veit,
R.~R. (1994). Partial differential equations in ecology: spatial
interactions and populations dynamics. \emph{Ecology} \textbf{75},
pp. 17-29.

\bibitem{HURFORD} Hurford, J.~R., (2003). Why synonymy is rare:
Fitness is in the speaker. In Banzhaf, W., Christaller, T.,
Dittrich, P., Kim, J.~T., and Ziegler, J. (eds.), \emph{Advances
in artificial life - Proceedings of the 7th European Conference on
Artificial Life (ECAL), lecture notes in artificial intelligence},
vol. \textbf{2801}, pp. 442-451 (Berlin: Springer Verlag).

\bibitem{JACKENDOFF} Jackendoff, R.~S., (1992). \emph{Languages of the mind}, (MIT Press).

\bibitem{KHIBNIK1997} Khibnik and Kondrashov (1997), Three
mechanisms of the Red Queen dynamics, \emph{Proc.~R.~Soc.~Lond.~B}
\textbf{264}, pp. 1049-1056.

\bibitem{KIRBY2002} Kirby, S., (2002). Natural Language from Artificial Life.
\emph{Artif.~Life} {\bf 8}, pp. 185--215.

\bibitem{KIRBY2001} Kirby, S. and Hurford, J., (2001). {\it The emergence of Linguistic
Structure; An Overview of the Iterated Learning Model}, In {\it
Simulating the Evolution of Language}, Cangelosi, A. and Parisi,
D. (eds.) (Springer-Verlag, Berlin).

\bibitem{KNIGHT} Knight, C., et al. (eds.) (2000).
\emph{The Evolutionary Emergence of Language Social Function and
the Origin of Linguistic Form}, (Cambridge University Press).

\bibitem{KOWALIK} Kowalik, M., Lipowski, A., and Ferreira,
A.~L.~(2002). Oscillations and dynamics in a two-dimensional
prey-predator system. \emph{Phys.~Rev.~E} \textbf{66}, pp.
066107-1--066107-5.

\bibitem{KOWALIKPHD} Kowalik, M. (2003). Badanie oscylacji
czasowych w uk\l adach makro\-sko\-po\-wych. PhD. thesis, Adam
Mickiewicz University, Pozna\'n Poland (in Polish).

\bibitem{LIP1999} Lipowski, A. (1999). Oscillatory behaviour in a lattice
prey-predator system. \emph{Phys. Rev. E} \textbf{60}, pp.
5179--5184.

\bibitem{LIP2000} Lipowski, A., (2000). Multiplicity of species in some replicative
systems, \emph{Phys. Rev. E} \textbf{61}, pp. 3009--3014.

\bibitem{LIP2005} Lipowski, A. (2005). Periodicity of mass extinctions without an
extraterrestrial cause. \emph{Phys. Rev. E} \textbf{71}, pp.
052902--052905.

\bibitem{LIPDROZ} Lipowski A. and Droz, M. (2002). Phase transitions in nonequilibrium d-dimensional models with q
absorbing states. \emph{Phys. Rev. E} \textbf{65}, pp. 056114-1 --
056114-7.

\bibitem{LIPLIP2000} Lipowski A. and Lipowska, D. (2000). Nonequilibrium phase transition
in a prey-predator system. \emph{Physica A} \textbf{276}, pp.
456--464.

\bibitem{LIPLIP2006} Lipowski A. and Lipowska, D. (2006). Long-term evolution of an
ecosystem with spontaneous periodicity of mass extinctions.
\emph{Theory in Biosciences} \textbf{125}, pp. 67--77.

\bibitem{LIPLIP2008} A.~Lipowski and Lipowska, D. (2008). Bio-linguistic transition
and the Baldwin effect in the evolutionary naming game model.
\emph{Int.~J.~Mod.~Phys.~C} \textbf{19}, pp. 399--407.

\bibitem{LIPLIP2008A} A.~Lipowski and Lipowska, D. (2008). Computational approach
to the emergence and evolution of language - evolutionary naming
game model. E-print: arXiv:0801.1658.

\bibitem{LIPLIP2008B} A.~Lipowski and Lipowska, D. (2008). Homonyms and synonyms in the \mbox{n--object} naming game model.
E-print: arXiv:0810.3442.

\bibitem{LOTKA} Lotka, A.~J.~(1920).
Analytical Note on Certain Rhythmic Relations in Organic Systems.
\emph{Proc. Natl. Acad. Sci. USA} \textbf{6}, pp. 410--415.

\bibitem{MATSUDA} Matsuda, H., Ogita, N., Sasaki, A., and Sato, K. (1992).
Statistical mechanics of population: the Lotka-Volterra model.
\emph{Prog.~Theor.~Phys.}~\textbf{88}, pp. 1035-1049.

\bibitem{MAY72} May, R. (1972). Limit Cycles in Predator-Prey
Communities. \emph{Science} \textbf{177}, pp.~900-902.

\bibitem{MSMITH1997} Maynard-Smith, J. and Szathm\'ary, E. (1997).
\emph{The Major Transitions in Evolution} (Oxford University
Press, New York).

\bibitem{MILLER} Miller, A.~I. and Foote, M., (2003). Increased Longevities of
Post-Paleozoic Marine Genera After Mass Extinctions.
\emph{Science} \textbf{302}, pp. 1030--1032.

\bibitem{MOBILIA} Mobilia, M., Georgiev, I.~T., and Ta\"uber U.~C.
(2005). Fluctuations and Correlations in Lattice Models for
Predator-Prey Interaction. \emph{Phys.~Rev.~E} \textbf{73}, p.
040903.

\bibitem{MONETTI} Monetti, R., Rozenfeld, A.~F.~and Albano, E.~F. (2000).
Study of interacting particle systems: the transition to the
oscillatory behavior of a prey-predator model. \emph{Physica A}
\textbf{283}, pp. 52--58.

\bibitem{MUNROE2002} Munroe, S. and Cangelosi, A. (2002).
Learning and the evolution of language: the role of cultural
variation and learning cost in the Baldwin Effect. Artif.~Life
{\bf 8}, pp. 311-339.

\bibitem{MURRAY} Murray, J.~D.~(2002). \emph{Mathematical Biology Vols.
I/II} (Springer-Verlag, New York).

\bibitem{NETTLE1999} Nettle, D., (1999). Using Social Impact Theory to simulate language change.
\emph{Lingua} {\bf 108}, pp. 95--117.

\bibitem{NETTLE1999A} Nettle, D., (1999). Is the rate of linguistic change constant?
\emph{Lingua} {\bf 108}, pp. 119--136.

\bibitem{NEWMANPALMER2003} Newman, M.~E.~J.~and Palmer, R.~G.
(2003). \emph{Modelling Extinction} (Oxford University Press, New
York).

\bibitem{NOBLE} Noble, J., (2000). In \emph{Evolutionary Emergence of Language},
Knight, C. et al. (eds.) (Cambridge University Press).

\bibitem{NOWAK} Nowak, M.~A. and Komarova, N.~L., (2001). Towards
an evolutionary theory of language. \emph{Trends in
Cogn.~Sci.}~{\bf 5}, pp. 288--295.

\bibitem{NOWAK1999} Nowak, M.~A. and Krakauer, D.~C., (1999). The evolution of language.
\emph{Proc.~Natl.~Acad.~Sci.~USA} {\bf 96}, pp. 8028--8033.

\bibitem{NOWAK2002} Nowak, M.~A., Komarova, N.~L., and Niyogi, P. (2002).
Computational and evolutionary aspects of language.
\emph{Nature}~{\bf 417}, pp. 611--617.

\bibitem{PAULO2007} de Oliveira, P.~M.~C., Stauffer, D., Wichmann, S., and de
Oliveira, S.~M., (2007). A computer simulation of language
families. E-print:arXiv:0709.0868.

\bibitem{ORGEL} Orgel, L.~E., (1992). Molecular replication, \emph{Nature} \textbf{358},
pp. 203--209.

\bibitem{PATRIARCA} Patriarca, M. and Heinsalu, E. (2008).
Influence of geography on language evolution. \emph{Physica A} (in
press).

\bibitem{PATTERSON} Patterson, C. and Smith, A.~B. (1989). Periodicity in extinction:
the role of the systematics. \emph{Ecology} \textbf{70}, pp.
802--811.

\bibitem{PEKALSKI} P\c{e}kalski, A. (2004).
A short guide to predator-prey lattice models. \emph{Computing in
Science and Engineering} \textbf{6}, pp. 62--66.

\bibitem{PINKER1990} Pinker, S. and Bloom, P., (1990). Natural language and natural selection.
\emph{Behav.~Brain Sci.}~{\bf 13}, pp. 707--784.

\bibitem{PLOTNICK1993} Plotnick, R.~E., and McKinney, M.~L., (1993). Ecosystem organization
and extinction dynamics. \emph{Palaios} \textbf{8}, pp.~202--212.

\bibitem{PLOTNICK} Plotnick, R.~E. and Sepkoski, J.~J., (2001). A multiplicative
multifractal model for originations and extinctions.
\emph{Paleobiology} \textbf{27}, pp. 126--139.

\bibitem{PROKOPH} Prokoph, A., Fowler, A.~D., and Patterson, R.~T., (2000). Evidence
for periodicity and nonlinearity in a high-resolution fossil
record of long-term evolution. \emph{Geology} \textbf{28}, pp.
867--870.

\bibitem{PROVATA} Provata, A., Nicolis, G., and Baras, F. (1999).
Ocillatory Dynamics in Low Dimensional Lattices: A Lattice
Lotka-Volterra Model. \emph{J.~Chem.~Phys.}~\textbf{110}, pp.
8361--8368.

\bibitem{RAI} Rai, R. and Singh, H.~(2000).
Stochastic resonance without an external periodic drive in a
simple prey-predator model. \emph{Phys.~Rev.~E} {\bf 62}, pp.
8804--8807.

\bibitem{RAMPINO} Rampino, M.~R. and Stothers, R.~B., (1984). Terrestrial mass
extinctions, cometary impacts and the Sun's motion perpendicular
to the galactic plane. \emph{Nature} \textbf{308}, pp. 709--712.

\bibitem{RAUPSEPKOSKI1984} Raup, D.~M.~and Sepkoski, J.~J. (1984).
Periodicities of extinctions in the geologic past.
\emph{Proc.~Natl.~Acad.~Sci.~U.S.A.~}{\bf 81} pp. 801--805.

\bibitem{ROHDE} Rohde, R.~A. and Muller, R.~A., (2005). Cycles in fossil
diversity. \emph{Nature} \textbf{434}, pp. 208--210.

\bibitem{ROSENFELD} Rozenfeld, A.~F. and Albano, E.~V.~(2001).
Critical and oscillatory behavior of a system of smart preys and
predators. \emph{Phys. Rev. E} \textbf{63}, pp. 061907.

\bibitem{SAMPSON} Sampson, G., (1997). {\it Educating Eve: The 'Language Instinct' Debate},
(Cassell).

\bibitem{SATULOVSKY} Satulovsky, J.~E. and Tom\'e, T. (1994).
Stochastic lattice gas model for a predator-prey system.
\emph{Phys.~Rev.~E} \textbf{49}, pp. 5073--5079.

\bibitem{SCHAFFER} Schaaffer, W. (1984). Stretching and folding in
lynx fur returns: Evidence for a strange attractor in nature.
\emph{Am.~Nat.}~\textbf{124}, pp. 798-820.

\bibitem{STAUFFER1} Stauffer, D. (1977). Percolation clusters as
teaching aid for Monte Carlo simulation and critical exponents.
\emph{Am.~J.~Phys}. \textbf{45}, pp. 1001.

\bibitem{STAUFFER2} Stauffer, D. (1979). Scaling theory of
percolation clusters. \emph{Phys.~Rep}.~\textbf{54}, pp. 1.

\bibitem{STAUFFER2008} Schulze, C., Stauffer, D., and Wichmann, S., (2008).
Birth, survival and death of languages by Monte Carlo simulation.
Commun.~Comp.~Phys.~ {\bf 3}, pp. 271-294.

\bibitem{SIMPSON1983} Simpson, G.~G. (1983). \emph{Fossils and the history of
life}(Scientific American Library, New York).

\bibitem{SIMPSON} Simpson, G.~G., (1953). The Baldwin Effect. \emph{Evolution} {\bf 7}, pp. 110--117.

\bibitem{SMITH} Smith, K., (2003)
The Transmission of Language: models of biological and cultural
evolution. Ph.D.~thesis, (The University of Edinburgh).

\bibitem{SOLE} Sol\'e R.~V. and Manrubia, S.~C. (1996). Extinction
and self-organized criticality in a model of large-scale
evolution. \emph{Phys.~Rev. E} \textbf{54}, pp. R42-R45.

\bibitem{STANLEY1973} Stanley, S.~M, (1973). An explanation for Cope's rule. \emph{Evolution}
\textbf{27}, pp.~1--26.

\bibitem{STANLEY1990} Stanley, S.~M, (1990). Delayed recovery and the spacing of major
extinctions. \emph{Paleobiology} \textbf{16}, pp.~401--414.

\bibitem{STEELS1995} Steels, L. (1995). A self-organizing spatial vocabulary.
\emph{Artif.~Life} {\bf 2}, pp. 319--332.

\bibitem{STEELS2002} Steels, L., (2002). In {\it Proceedings of the International
Workshop of the Self-Organization and Evolution of Social
Behaviour}, Hemelrijk, C. and Bonabeau, E. (eds.) (University of
Zurich, Switzerland).

\bibitem{STOTHERS1986} Stothers, R.~B., (1986). Periodicity of the Earth's magnetic
reversals. \emph{Nature} \textbf{322}, pp. 444--446.

\bibitem{STOTHERS1993} Stothers, R.~B., (1993). Flood basalts and extinction events,
\emph{Geophys. Res. Lett.} \textbf{20}, pp. 1399--1402.

\bibitem{SZAMADO} Sz\'amad\'o, S. and Szathm\'ary, E., (2006). Language Evolution: Competing
Selective Scenarios. \emph{Trends.~Ecol.~Evol.}~{\bf 21}, pp. 555.

\bibitem{SZATHMARY} Szathm\'ary, E., (1999). The origin of the genetic code. \emph{Trends in
Genetics} \textbf{15}, pp. 223--229.

\bibitem{THOMPSON} Thompson, J.~N. (1998). Rapid evolution as an ecological process.
\emph{Trends in Ecol.~Evol.}~\textbf{13}, pp. 329--332.

\bibitem{TRIVERS} Trivers, R.~L., (1971).The evolution of reciprocal altruism.
\emph{Quart.~Rev.~Biol.} {\bf 46}, pp. 35--57.

\bibitem{TURNEY1996} Turney, P., (1996). {\it Myths and legends of the Baldwin Effect}.
In Fogarty, T. and Venturini, G. (Eds.), Proceedings of the
ICML-96 (13th International Conference on Machine Learning, Bari,
Italy).

\bibitem{VANDEVALLE} Vandevalle, N. and Ausloos, M. (1995). The
robustness of self-organized criticality against extinctions in a
tree-like model of evolution. \emph{Europhys.~Lett.} \textbf{32},
pp. 613-618.

\bibitem{VANVALKEN} Van Valkenburgh, B., Wang, X., and Damuth, J., (2004). Cope's Rule,
Hypercarnivory, and Extinction in North American Canids.
\emph{Science} \textbf{306}, pp.~101--104.

\bibitem{VERHULST} Verhulst, P.~F.~(1838).
Notice sur la loi que la population poursuit dans son
accroissement. \emph{Corresp. Math. Phys.} \textbf{10},
pp.113--121.

\bibitem{VOLTERRA} Volterra, V.~(1926). Variazioni e fluttuazioni
del numero d'individui in specie animali conviventi. \emph{Mem.
Accad. Nazionale Lincei}, ser.~6, \textbf{2}, pp. 31-112.

\bibitem{WEBER} Weber, B.~H. and Depew, D.~J. (eds.) (2003).
{\it Evolution and Learning - The Baldwin Effect Reconsidered},
(Cambridge, MA, MIT Press).

\bibitem{YAMA2004} Yamauchi, H. (2004). \emph{Baldwinian Accounts of Language
Evolution}. Ph.D. thesis, University of Edinburgh, Edinburgh,
Scotland.
%%%%%%%%%%%%%%%%%%%%%%%%%%%%%%%%%%%%%%%%%%%%%%%%%%%%%%%%%%%%%
\end{thebibliography}
\end{document}